# Evaluation of No Reference Bitstream-based Video Quality Assessment Methods

Tiantian He[*], Yankai Liu[*], Rong Xie[†], Xin Tang[‡], Li Song[†]
[*]Institute of Image Communication and Network Engineering, Shanghai Jiao Tong University, Shanghai, China
[†]Future Medianet Innovation Center, Shanghai, China
[‡]Video Solution dept, Huawei technologies co., ltd, Shenzhen, China
E-mail: {hetiantian, liuyankai, xierong}@sjtu.edu.cn, Donald.tangxin@huawei.com, song_li@sjtu.edu.cn

*Abstract*—Many different parametric models for video quality assessment have been proposed in the past few years. This paper presents a review of nine recent models which cover a wide range of methodologies and have been validated for estimating video quality due to different degradation factors. Each model is briefly described with key algorithms and relevant parametric formulas. The generalization capability of each model to estimate video quality in real-application scenarios is evaluated and compared with other models, using a dataset created with video sequences from practical applications. These video sequences cover a wide range of possible realistic encoding parameters, labeled with mean opinion scores (MOS) via subjective test. The weakness and strength of each model are remarked. Finally, future work towards a more general parametric model that could apply for a wider range of applications is discussed.

## I. INTRODUCTION

Video quality assessment (VQA) has been greatly promoted these days by the proliferation of video applications, including video on demand, videoconference, videophone and so on. It is crucial to guarantee a satisfactory quality of experience (QoE) from the user perspective, and video quality is one of the most direct and critical factors in the user QoE. Many artifacts can lead to quality degradations when the video is processed and transmitted. The aim of video quality assessment is to evaluate the perceptual quality of video with all kinds of degradations.

Video quality assessment methods can be divided into two main categories: subjective assessment methods and objective assessment methods. A subjective test is conducted in a standardized environment, using a set of video sequences processed by different options to measure people's mean opinion scores (MOS) [1]. Subjective tests are the most reliable but cumbersome, time consuming and expensive methods. On the contrary, objective assessment methods, which use algorithms to predict the perceived video quality in different scenarios, are both efficient and effective. According to the amount of reference information about the original video, objective video quality assessment can be further classified into three different types: full-reference (FR), reduced reference (RF) and no reference (NR) models. NR models make the assessment of perceived video quality only based on the degraded video sequences without any reference. In many practical application scenarios, the original video is unavailable and only NR models are applicable under the circumstances, thus the research on NR models is of great practical significance.

According to the type of information used in the evaluation process, NR models can be classified into bitstream layer models, packet layer models and hybrid models. In bitstream layer models, the assessment operates on the parsed coding parameters from bitstream (e.g. quantization parameter, frame rate, bit rate, motion vector). Packet layer models only use general packet header information about the network (e.g. packet loss rates), and do not take into account media related information. Hybrid models exploit both the packet header information and bitstream parameters. Detailed instruction about video quality assessment methods is illustrated in Fig.1. The highlighted parts belong to the research scope of this paper.

Parametric models access the perceived video quality based on a set of extracted parameters which could be fundamental factors leading to the video quality degradations. Typical video degradations include video compression artifacts and packet loss yielding frame loss, slicing and freezing artifacts. In this paper, we mainly focus on the degradation types relevant to bitstream layer regardless of other degradation types in packet layer (i.e. packet loss).

Over the years, many groups and researchers have made significant contributions towards parametric NR video quality assessment, including the standardized work by International Telecommunication Union (ITU-T) and other published models of previous researchers. Focusing on particular scopes or applicability in certain scenarios (e.g. particular video resolution), each model has been typically designed and separately validated by their authors. But in real-world scenarios, video contents and coding parameters could be various and non-standard. Under the circumstances, the performances of these models, which have been separately evaluated on specifically encoded (i.e. using standardized encoding parameters) video sequences, could be degraded with non-stable predictions. Previous researchers have made reviews and comparisons of some models. Yankai Liu et al. [2] have provided a review of standardized ITU-T NR parametric models for compressed video quality estimation, including ITU-T Rec. G.1070 [3], ITU-T Rec.P.1201 series [4], ITU-T Rec. P.1202 series [5] and ITU-T Rec. J.343.1 [6]. The performances of these models were detailed compared in their work, however only standardized efforts were included and the comparison was based on the separately evaluated results of each model. Joskowicz et.al [7] compared the performances of ten NR parametric models for perceptual video quality, but only one standardized model Rec. G.1070 was included in their work and the dataset used in their comparison belongs to standardized encoding work rather than the uncertain practical encoding work. Up to the time of this paper, there is no work analyzing ITU-T P.1203 [8] and there lacks a comprehensive evaluation of generalization capability of both standardized and other published models with real-world testing data.

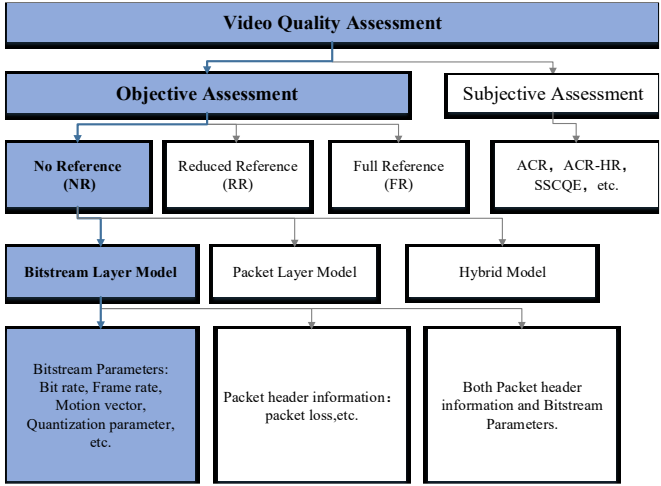

Fig.1 Classification of video quality assessment.
(The highlighted parts indicate the research direction of this paper.)

In this paper, we present a review of nine parametric models proposed by different groups of authors, including four standardized models by ITU-T, other four models by previous researchers and one model adopted in *China Video Service Experience* by China Video Service User Experience Standard Working Group [17]. The performances of each model to estimate video degraded quality were evaluated on a practical dataset, and the strength and weakness of these models were vividly remarked. Based on the comparison results, we offered a discussion about the development of a more general parametric model for video quality assessment.

The remainder of the paper is organized as follows: Section II provides a brief review of nine previous proposed parametric VQA models. In Section III, experimental results are provided and performances of these models are statistically compared and discussed. Section IV summaries the results and discusses some future research issues.

## II. PARAMETRIC MODELS

In this section, nine previous models, which cover a wide range of methodologies, are reviewed briefly. These models may be capable of evaluating many degradation artifacts like compression artifacts, re-buffering artifacts and packet-loss artifacts. Since we only focus on the bitstream parameters in this paper, degradation artifacts due to packet information in these models (if any) are not taken into consideration in this review.

### A. ITU-T Rec. G.1070

This recommendation describes a computational model to assess video quality for point-to-point interactive videophone application areas [3]. The effect of video content is not taken into account in this model. Video quality $QV$ is calculated based on coding distortion $I_{coding}$ as:

$$QV = 1 + I_{coding} \quad (1)$$

$$I_{coding} = I_{ofr} \exp\left\{-\frac{\left(\ln(fr_v) - \ln(O_{fr})\right)^2}{2 D_{fr_v}^2}\right\} \quad (2)$$

where $fr_v$ denotes video frame rate, $I_{ofr}$ is the maximum video quality at each bit rate, calculated as (3), $O_{fr}$ is the optimal frame rate that maximizes the video quality at each bit rate $br_v$ as (4), and $Dfr_v$ is the degree of video quality robustness due to frame rate as (5).

$$I_{ofr} = a_1 - \frac{a_2}{1 + (br_v / a_3)^{a_4}}, 0 \leq I_{ofr} \leq 4 \quad (3)$$

$$O_{fr} = a_5 + a_6 \cdot br_v, 0 \leq O_{fr} \leq 30 \quad (4)$$

$$Dfr_v = a_7 + a_8 \cdot br_v \quad (5)$$

In equations above, $a_1, a_2, \ldots$ and $a_8$ are coefficients of the model.

### B. ITU-T Rec. P.1201.1

The standardized Recommendation ITU-T Rec. P.1201.1 [9] is one of the algorithmic models described in Recommendation ITU-T P.1201-series [4], and another one is ITU-T Rec. P.1201.2 [10]. This recommendation addresses the lower resolution application area to monitor the performance and QoE of video services, such as Mobile TV. In this model, $fr_v$ and $br_v$ denote video frame rate and bit rate respectively. Video quality due to compression $QV$ is calculated as (6) when the video frame rate is less than 24, as (7) otherwise.

$$QV = (5 - Qcod) \quad (6)$$

$$QV = (5 - Qcod)\left(1 + c_1 \cdot cpx_{video} - c_2 \cdot cpx_{video} \cdot \log\left(\frac{1000}{fr_v}\right)\right) \quad (7)$$

where $Qcod$, calculated as (8), denotes the video distortion quality by compression artifacts, $normbr_v$ is the normalized video bit rate calculated as (9) and $cpx_{video}$ denotes the video content complexity factor, calculated as (10).

$$Qcod = \frac{4}{1 + \left(\frac{normbr_v}{c_3 \cdot cpx_{video} + c_4}\right)^{(c_3 \cdot cpx_{video} + c_4)}} \quad (8)$$

$$normbr_v = \frac{br_v \cdot 8 \cdot 30}{1000 \cdot \min(30, fr_v)} \quad (9)$$

$$cpx_{video} = \min\left(\sqrt{\frac{br_v}{AvgByte_{I\text{-}frame}}}, 1.0\right) \quad (10)$$

In equations above, $cpx_{video}$ describes the spatio-temporal complexity of video content with a maximum value 1.0 and initial value 0.5, $AvgByte_{I\text{-}frame}$ represents the average number of bytes per I-frame and $c_1$, $c_2$, $c_3$, $c_4$ are coefficients of this model.

### C. ITU-T Rec. P.1201.2

Complementary to the ITU-T P.1201.1, ITU-T Rec. P.1201.2 [10] specifies algorithms for the higher resolution application areas, such as Internet Protocol television (IPTV). Regardless of the packet loss degradation, the predicted video quality $QV$ is decomposed as:

$$QV = 100 - Qcod \quad (11)$$

where $Qcod$ denotes the video distortion quality by compression artifacts, calculated as:

$$Qcod = c_1 \cdot \exp(c_2 \cdot bitPerPixel) + c_3 \cdot cpx_{video} + c_4 \quad (12)$$

$$bitPerPixel = \frac{br_v \cdot 10^6}{numPixels_{frame} \cdot fr_v} \quad (13)$$

where $bitPerPixcel$ denotes the averaged bits per pixel in the video sequence, as calculate in (13) where $br_v$ is video bit rate, $fr_v$ is video frame rate and $numPixels_{frame}$ is the number of pixels per frame. Additionally, the content complexity parameter $cpx_{video}$ is calculated to estimate the spatio-temporal complexity of video content. It is given by:

$$cpx_{video} = \frac{\sum_{sc=1}^{Z} w_{sc} \cdot N_{sc}}{\sum_{sc=1}^{Z} S^I_{sc} \cdot w_{sc} \cdot N_{sc}} \cdot \frac{numPixels_{frame} \cdot fr_v}{1000} \quad (14)$$

The newly introduced variables in (14) are related to scene detection technology. Let $sc$ denotes a scene, then $N_{sc}$ is the number of Groups Of Pictures (GOPs) in it, $S^I_{sc}$ is the averaged I frame size in this sense and if this scene has the lowest $S^I_{sc}$, the value of $w_{sc}$ is 16, otherwise is 1. Limited to the space, detailed introduction about the scene detection technology is not provided here. Interested reader can refer to [11].

In equations above, $c_1$, $c_2$, $c_3$ and $c_4$ are coefficients.

### D. ITU-T Rec. P.1203.1 Mode 3

The ITU-T P.1203-series [8] describes a set of objective parametric bitstream-based modules to predict video and audio perceived quality of progressive download and adaptive audiovisual streaming services. Among these models, ITU-T Rec. P.1203.1 [12] describes the video quality assessment model, which is further classified into four modes (Mode 0, 1, 2, 3) restricted to the input information provided. Mode 3 is the most complex one that can operate with no encryption metadata and any information from video stream. This model provides video coding quality per output sampling interval calculated as the integration of three aspects: quantization degradation $Dq$ in (15), upscaling degradation $Du$ in (19) and temporal degradation $Dt$ in (22).

$$Dq = 100 - RfromMOS(MOSq) \quad (15)$$

$$Dq = \max(\min(Dq, 100), 0) \quad (16)$$

where $RfromMOS$ and $MOSfromR$ in (30) are equivalent transformation functions introduced in Annex E of [12].

$$MOSq = q_1 + q_2 \cdot \exp(q_3 \cdot quant) \quad (17)$$

$$MOSq = \max(\min(MOSq, 5), 1) \quad (18)$$

Here $quant$, whose computation is different between four modes, is the parameter that estimates the quantization degradation. It is computed based on a set of parameters including frame type, averaged QP and skip ratio of macroblocks per frame. With the same consideration of space, computation of $quant$ in Annex D of [12] is not detailed here.

Upscaling degradation is related with terminal display resolution $disRes$ and video resolution $codRes$. Let $scaleFactor$ denote the parameter capturing the upscaling degradation and the calculation is given by:

$$Du = u_1 \cdot \log_{10}(u_2 \cdot (scaleFactor - 1) + 1) \quad (19)$$

$$Du = \max(\min(Du, 100), 0) \quad (20)$$

$$scaleFactor = \max\left(\frac{disRes}{codRes}, 1\right) \quad (21)$$

The overall temporal degradation is combined efforts of three types: the pure temporal degradation in (24), the coding impact relevant in (25) and the spatial scaling relevant in (26).

$$Dt = \begin{cases} Dt_1 - Dt_2 - Dt_3, & framerate < 24 \\ 0, & framerate \geq 24 \end{cases} \quad (22)$$

$$Dt = \max(\min(Du, 100), 0) \quad (23)$$

$$Dt_1 = \frac{100 \cdot (t_1 - t_2 \cdot framerate)}{t_3 + framerate} \quad (24)$$

$$Dt_2 = \frac{Dq \cdot (t_1 - t_2 \cdot framerate)}{t_3 + framerate} \quad (25)$$

$$Dt_3 = \frac{Du \cdot (t_1 - t_2 \cdot framerate)}{t_3 + framerate} \quad (26)$$

The final degradation $D$ is calculated as follows:

$$D = \max(\min(Dq + Du + Dt, 100), 0) \quad (27)$$

Finally, quality $Q$ and $MOS$ are the estimated video encoding qualities with different scales.

$$Q = 100 - D \quad (28)$$

$$MOS = \begin{cases} MOSq, \text{if } Du = 0 \text{ and } Dt = 0 \\ MOSfromR(Q), \text{otherwise} \end{cases} \quad (29)$$

where $MOS \in [1:5]$ and $Q \in [0:100]$. This quality is calculated for the device type "TV". An adjustment on the final result is needed if the device type is "handheld":

$$MOSq_{handheld} = h_1 + h_2 \cdot MOSq + h_3 \cdot MOSq^2 + h_4 \cdot MOSq^3 \quad (30)$$

$$MOSq = \max(\min(MOSq_{handheld}, 5), 1) \quad (31)$$

In equations above, $q_1$, $q_2$, $q_3$, $u_1$, $u_2$, $t_1$, $t_2$, $t_3$, $h_1$, $h_2$, $h_3$ and $h_4$ are coefficients of this model.

### E. Yamagishi model

Yamagishi et al. [13] have established a function to estimate video quality effected by coding degradation. This function is based on the relationship between video quality, bit rate $br_v$ and frame rate $fr_v$. An optimal frame rate $f_O$ is used to maximum video quality value $MOS_p$ at each bite rate. The function of estimating coding distorted video quality $QV$ is approximated by a convex Gaussian function $v_c$:

$$QV = 1 + v_c \quad (32)$$

$$v_c = v_0 \cdot \exp\left(-\frac{(\ln(fr_v) - \ln(f_O))^2}{2D_{Fr}^2}\right) \quad (33)$$

$$f_O = c_1 + c_2 \cdot br_v \quad (34)$$

$$D_{Fr} = c_3 + c_4 \cdot br_v \quad (35)$$

$$v_0 = c_5 \left(1 - \frac{1}{1 + \left(\frac{br_v}{c_6}\right)^{c_7}}\right) \quad (36)$$

When $fr_v = f_O$, the value of $MOS_p$ indicates the optimal video quality, and $D_{Fr}$ indicates the frame rate related video quality degradation, and $v_0$ is the logistic relationship between $MOS_p$ and $br_v$. In equations above, $c_1$, $c_2$, ... and $c_7$ are coefficients of this model.

### F. Rises model

Rises et al. [14] have designed a model for low resolution video typical in video stream applications. They presented an algorithm to recognize the content type and identify the content into five classes. Based on the content class, video frame rate $fr_v$ and bitrate $br_v$, the assessment is performed to obtain the video quality $QV$ as:

$$QV = c_1 + c_2 br_v + \frac{c_3}{br_v} + c_4 fr_v + \frac{c_5}{fr_v} \quad (37)$$

where $c_1$, $c_2$, $c_3$, $c_4$ and $c_5$ are coefficients of this model.

### G. Joskowicz model

Joskowicz et al. [15] have developed a parametric model to evaluate the combined effects of bit rate $br_v$, and video content. The video quality estimation $QV$ is calculated according to the degradation degree $v_c$ due to coding as:

$$QV = 1 + v_c \quad (38)$$

$$v_c = 4\left(1 - 1 / \left(1 + \left(\frac{br_v}{v_1}\right)^{v_2}\right)\right) \quad (39)$$

where $v_1$ and $v_2$ are model coefficients related to average Sum of Average Differences (SAD) per pixel of the clip. SAD provides an overall assessment of the spatial-temporal activity such as moving vectors calculations.

$$v_1 = c_1 SAD^{c_2} + c_3 \quad (40)$$

$$v_2 = c_4 SAD^{c_5} + c_6 \quad (41)$$

In equations above, $c_1$, $c_2$, ... and $c_6$ are coefficients of this model.

### H. Takagi model

Takagi et al. [16] have proposed a method to estimate video quality with arbitrary resolutions and frame rates. This model estimates the relationship between quantization parameter $QP$ and bit rate $br_v$ for further estimation of the coding complexity $v_c$ in the MOS assessment process.

$$MOS_p = \frac{-1}{\frac{1}{\gamma} + e^{(\alpha(QP-\beta))}} + \gamma \quad (42)$$

where $\alpha$, $\beta$ and $\gamma$ represent the dependency on video resolution $r$, frame rate $fr_v$ and video coding complexity $v_c$ as:

$$\alpha = a_1 \log(r) + b_1 \log(fr_v) + c_1 \log(v_c) + d_1 \quad (43)$$

$$\beta = a_2 \log(r) + b_2 \log(fr_v) + c_2 \ln(v_c) + d_2 \quad (44)$$

$$\gamma = a_3 \log(r) + b_3 \log(fr_v) + c_3 \log(v_c) + d_3 \quad (45)$$

$$v_c = QP - e \ln(br_v) \quad (46)$$

In equations above, $a_1$, $a_2$, $a_3$, $b_1$, $b_2$, $b_3$, $c_1$, $c_2$, $c_3$, $d_1$, $d_2$, $d_3$ and $e$ are coefficients of this model.

### I. uVES Mode 1

In 2017, *China Video Service Experience* is published [17] as the national industry standard in China. There are three video quality assessment models in this standard, among which uVES Mode 1 operates with the encoding parameters to estimate video quality on a particular device. It can be decomposed into two sub-models named Model 1.1 and Model

1.2 respectively. The final estimated video quality $Qs$ is calculated by the results of two sub-models:

$$Qs = Qd - (5 - Qcod)\frac{Qd - N_1}{N_2} \quad (47)$$

where Qcod represents encoding quality in Model 1.1, Qd represents display quality in Model 1.2, N1 = 4 and N2 = 100.

*1) Model 1.1*

Model 1.1, which aims at the encoding distortion of video quality, is a modification based on ITU-T Rec. P.1202.1. The parametric formulas of Model 1.1 operate by analyzing impacts of quantization $OP\_fr$, motion vector $MV_{imp}$, video complexity $cpx_{video}$ and key frame rate indicator $kfr_{imp}$ as:

$$Qcod = kfr_{imp} \cdot \exp\left(n_1 \cdot (QP\_fr + cpx_{video} + MV_{imp})\right) \quad (48)$$

$$Qcod = \min(\max(Qcod,1),5) \quad (49)$$

$$kfr_{imp} = n_2 \cdot kfr + n_3 \quad (50)$$

Here $kfr$ denotes the key frame rate obtained by $kfr = fr / d$, where $fr$ denotes the average frame rate and $d$ denotes the average frame number between two intra frames.

$$QP\_fr = n_4 + n_5 \cdot \left(\frac{avgQP}{51}\right)^{n_6} + n_7 \cdot \frac{1}{fr} + n_8 \cdot Iflicker \\ + n_9 \cdot (maxQP - minQP) \quad (51)$$

Here $avgQP$, $maxQP$ and $minQP$ denote the averaged quantization parameter of all macroblocks, the maximum and minimum value when averaging macroblock QP per picture respectively. *Ifilcker* represents the intra-picture flicker to detect the abrupt change of averaged QP. The intra-picture flicker is identified when the differences of averaged QP between current and preceding picture, current and subsequent picture are both higher than a threshold value 5.

$$cpx_{video} = \min\left(\sqrt{\frac{br}{AvgByte_I}} + n_{10} \cdot skipRatio, 1\right) \quad (52)$$

The video content spatio-temporal complexity indicator is calculated in (53), where $AvgByte_I$ denotes the averaged bytes per I-frame and $br$ is video bit rate. Additional parameter used is the ratio of skipped macroblocks per frame, introduced as $skipRatio$.

$$MV_{imp} = n_{11} \cdot avgMV \cdot \left(1 - \frac{fr}{30}\right) \quad (53)$$

where $avgMV$ is the absolute vertical and horizontal motion vectors averaged over macroblocks in the sequence.

*2) Model 1.2*

Model 1.2 focuses on the display quality based on extracted coding parameter $avgQP$ as well as display parameters. $video\_width$ and $video\_height$ denote the width and height of video in pixels; $screen\_size$ denotes screen size of display devices in inches and $ppi$ denotes pixels per inch. The display quality Qd is calculated as:

$$Qd = Q_{Disp} - \frac{(Q_{Disp} - 1)}{1 + \exp\left(\frac{n_{12} - avgQP}{n_{13}}\right)} \quad (54)$$

$$Q_{disp} = n_{14} \cdot \left(1 - 1 \div \left(1 + \left(\frac{ppi}{n_{15} \cdot (screen\_size)^{n_{16}}}\right)^{n_{17}}\right)\right) \quad (55)$$

$$Q_{Disp} = \max\left(1, \min\left(5, Q_{disp}\right)\right) \quad (56)$$

$$ppi = \frac{\sqrt{video\_width^2 + video\_height^2}}{screen\_size} \quad (57)$$

In equations above, $n_1, n_2, ...$ and $n_{17}$ are the coefficients of this model.

III. COMPARISON AND DISCUSSION

As we can see from the previous sections, many parametric models have been proposed, taking into account some specific parameters (i.e. frame rate, bit rate, video content and so on) and specific application conditions. Table I provides a comprehensive summary of these reviewed models in this paper. This section will present the comparison results of different models and discussions of the development of a more general parametric model.

To fairly evaluate the performance of parametric video quality assessment models reviewed in Section II, the evaluation of different model is conducted on the same dataset. Our experiment aims at evaluating the applicability and robustness of these models to estimate the quality of video with arbitrary compression parameters in real-application scenarios.

This dataset is provided by China Video Service User Experience Standard Working Group, and the video sequences are collected from real-applications by the members of the working group, including China Unicom, China Telecom, etc. This dataset contains 115 video sequences covering a wide range of possible realistic coding parameters without any packet layer degradation types. The video sequences in the dataset are encoded in H.264/AVC format using different profiles (main and high) with different content (e.g. anime, drama, science fiction, sports videos, documentaries, variety show and so on), different frame rate and bit rate, random GOP structure and size as shown in Table II. Each sequence in this dataset is labeled with a subjective MOS. The subjective MOS are tested under standard procedures in Recommendation ITU-T Rec.P910 [18], using ACR five-point numerical quality scale. We will soon publish this dataset.

We conduct the performance evaluation of each model by the k-fold cross validation. All the sequences are randomly partitioned into k equal subsets. Only one single subset is retained to test the model and the remaining k-1 subsets are used as the training data. The validation process repeats k times to ensure each one of the subset is used exactly once as the test data. The final single result is averaged over all the k-times validation results. K-fold cross validation is an effective method to evaluate the stability of the model, especially when the dataset is not in a very large scale. This method ensures that all data can be used both for training and testing models and the performance of model can be evaluated k times. Considering the computational expense and dataset scale, we choose five-fold cross validation in this experiment. The video sequences from different source are randomly partitioned and those from the same source are carefully split into different subsets.

TABLE I
MODEL COMPARISON

| Model | Equations | Bit rate | Frame rate | Video Content | Display parameters | Coef | Tested Conditions |
|---|---|---|---|---|---|---|---|
| ITU-T G1070 | 1-5 | Yes | Yes | No | No | 8 | VGAQ, VGA, QQVGA; MPEG4, H.264 |
| ITU-T P.1201.1 | 6-10 | Yes | Yes | Yes- Content Complexity | No | 4 | HVGA, QVGA, QCIF; MPEG4, H.264 |
| ITU-T P.1201.2 | 11-14 | Yes | Yes | Yes- Content Complexity (Scene Detection) | No | 4 | SD, HD; H.264 |
| ITU-T P.1203.1 Mode 3 | 15-31 | Yes | Yes | Yes- QP, Skip ratio | Yes-Device types, Display resolution | 12 | HD; H.264 |
| Yamagishi | 32-36 | Yes | Yes | No | No | 7 | VGA, QVGA, MPEG4 |
| Rises | 37 | Yes | Yes | Yes-Content Classes | No | 5 | CIF, QCIF, SIF; H.264 |
| Joskowicz | 38-41 | Yes | Yes | Yes-SAD | No | 6 | SD, VGA, QCIF, CIF; H.264 |
| Takagi | 42-46 | Yes | Yes | Yes- Content Complexity | No | 12 | HD, 1280×720, 960×540, 640×360, 480x270; H.264 |
| uVES Mode 1 | 47-57 | Yes | Yes | Yes- Content Complexity, Motion vector, QP, Skip ratio, intra-picture flicker | Yes- Screen size | 17 | QICF, QVGA, HVGA; H.264 |

TABLE II
DATASET PARAMETERS

| Resolution | 44 kinds of resolution from SD (416*176) to UHD (3840*2160) |
|---|---|
| Frame rate (fps) | 15,24,25,30 |
| Bit rate (kbps) | 180~8000 |
| Video codec and profile | H264/AVC-high and main |
| Frame duration | default |
| Segment duration | default |
| GOP structure and size | Random |
| QP | Random |
| Packet loss | None |

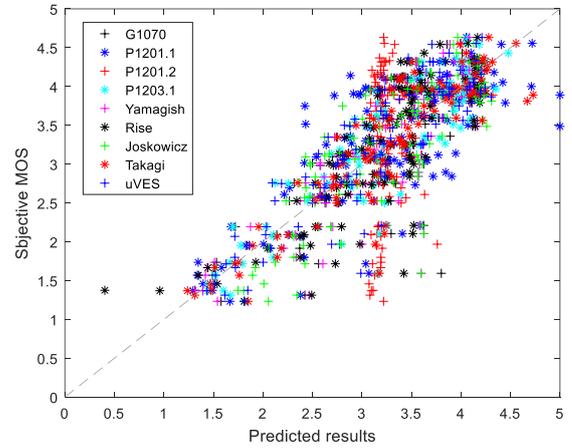

Fig.2 Scatter plot of subjective MOS versus predicted objective MOS by different models.

The performances of different models are compared in three main aspects as recommended in ITU-T Rec.P1401 [19]: we employ the Pearson Correlation Coefficient (PCC) to measure the linearity between estimated results and the subjective MOS, Root Mean Square Error (RMSE) to indicate the accuracy and outlier ratio (OR) to evaluate the consistency of the objective models. An outlier is defined using 95 per cent confidence interval as a threshold. Usually lager value of PCC, lower RMSE and lower outlier ratio mean a better performance.

The scatter plots of subjective MOS with respect to the predicted quality are illustrated in Fig.2. Note that all the results of five tests are plotted in one figure. The predicted results of different models are generally consistent with the subjective results. But some points show large prediction errors.

The performance validated with PCC, RMSE and outlier ratio averaged over all the five-fold cross validation is shown in Table III. In this comparison, the perfect performance is expected to have the highest PCC value 1, lowest RMSE value 0 and lowest OR value 0. The performances of some models evaluated in real-application scenarios are different from the experimental scenarios in their design papers.

*A. Evaluation of the dataset*

First we presented the evaluation of the database by a comparison between the recommended and re-trained best values of coefficients in P.1203.1 mode 3, as shown in Table IV. Most coefficients, except $q_2$, are of the same order of magnitude. But since the value $q_2$ is rather small, the influence is not great. The performance of P.1203.1 mode 3 using recommended coefficients is shown in Table V. As we can see, the performance is less good than, but still comparable to the retrained performance. These evaluation results indicate the validity of the dataset.

TABLE III
Model Performance Comparison vs Subjective Tests

| Models | PCC | RMSE | OR |
|---|---|---|---|
| G.1070 | 0.8422 | 0.2694 | 45.22% |
| P.1201.1 | 0.7411 | 0.3507 | 65.22% |
| P.1201.2 | 0.7651 | 0.9067 | 73.91% |
| P.1203.1 mode 1 | **0.8901** | **0.2513** | 43.48% |
| Yamagishi | 0.8351 | **0.2488** | 50.43% |
| Rises | 0.8248 | 0.2583 | 53.91% |
| Joskowicz | 0.8081 | 0.2725 | 57.39% |
| Takagi | 0.8813 | 0.2732 | **36.52%** |
| uVES Mode 1 | **0.8875** | 0.2835 | **39.13%** |

TABLE IV
Recommended vs Re-trained Best Value Of Coefficients In P.1203.1 Mode 1

| Coefficients | Recommended | Re-trained |
|---|---|---|
| $q_1$ | 4.66 | 4.077 |
| $q_2$ | -0.07 | -3.7111E-05 |
| $q_3$ | 4.06 | 19.8997 |
| $u_1$ | 72.61 | 40.5146 |
| $u_2$ | 0.32 | 0.8156 |
| $t_1$ | 30.98 | 28.0521 |
| $t_2$ | 1.29 | 1.3109 |
| $t_3$ | 64.65 | 29.108 |

*B. Comparison of the models*

Statistical evaluation metrics are shown in Table III with the best two results shown in bold. Of all the models, P.1203.1 mode 3 has the highest PCC 0.8901, which means the best linearity between estimated results with the subjective scores. The PCC of uVES Mode 1 is also top-ranked. Most models (except P.1201 series) have similar RMSE values, which are no higher than 0.3. Takagi, uVES Mode 1 and P.1203.1 mode 3 have lower value of outlier ratio.

The accuracy and consistency of P1201.2 are less good according to the values of RMSE and OR in Table III. This drop in performance is also visible in Fig. 3, where some points are further away from the correlation line. This performance drop may be explained by observing again the parameters used in this model. The perceived video quality has a high dependency on the video content complexity, which is captured by scene detection based on GOP length and structure.

In realistic applications like IP television (IPTV) or over-the-top(OTT) video services, the GOP length and structure of a video sequence is likely to be various, thus the performance of P1201.2 is possibly affected under this condition. Other models that are not GOP based have better performance.

*C. Discussion about a general parametric model*

Bit rate, frame rate, video content, display parameters and codec are all relevant to the perceived video quality estimation. Most model only take into account only a subset of these parameters. All the models reviewed in this paper take into account bit rate and frame rate. And whether video content is considered or not is not explicitly related to the evaluation results. The discussion about the effect of terminal display parameters in video quality assessment is presented below.

TABLE V
Performance Of P.1203.1 Mode 1 Using Different coefficients

| Coefficients | PCC | RMSE | OR |
|---|---|---|---|
| Recommended | 0.8540 | 0.3956 | 58.26% |
| Re-trained | 0.8901 | 0.2513 | 43.48% |

TABLE VI
Model Performance Comparison vs Subjective Tests

| | PCC | RMSE | OR |
|---|---|---|---|
| uVES Model 1.1 (without terminal display parameters) | 0.8155 | 0.3261 | 53.91% |
| uVES Mode 1 overall (with terminal display parameters) | 0.8875 | 0.2835 | 39.13% |

P.1203.1 mode 3 and uVES Mode 1 performed better in this evaluation. They are the only two models that take into account the terminal display parameters. To investigate the usage of introducing terminal display parameters into models, a test was conducted to evaluate the performance of overall uVES Mode 1 and uVES Model 1.1 separately, as shown in Table VI. With display parameters taken into account, the PCC of the overall model is 0.67 higher, which means more excellent correlation properties. The number of outliers is effectively reduced by 21.96% and RMSE is reduced by 0.0426. From a statistical point of view, display parameters do enhance the performance of this model. And from the practical point of view, taking into account display parameters is also reasonable.

According to section B, uVES Mode 1 and P1203.1Mode 3 are more robust to various real-application coding setups. In various video services, such as IPTV, OTT and adaptive streaming media services, a large number of videos are not standardly encoded. Meanwhile, more types of terminal playback devices can also effect the perceived video quality. As a result, the development of video quality evaluation models need to make appropriate adjustments (e.g. introducing more efficient parameters) to keep pace with real-world application scenarios. Newly proposed models P1203 and uVES have set an example for the future development of a more general video quality assessment model. They introduce the terminal display parameters into model to effectively enhance the generalization capability.

Other quality degradation factors that were not evaluated in this paper may also affect the video perceived quality, such as re-buffering, packet loss and so on. For a more general model, these factors should be evaluated.

IV. CONCLUSIONS

We have presented a review of some existing standardized and other published no reference parametric video quality assessment models in this work. The performance of each model was evaluated and compared with other models, using a dataset that was created from real-applications and was validated in this paper. From the experimental results, the Mode 3 proposed in ITU-T P.1203.1 and the uVES Mode 1 proposed by CCSA showed better performance than others. Towards a more general model, another comparison between taking into account the terminal display parameters or not was conducted. The results proved that display factors do have statistically significant to improve the model performance. Other factors (e.g. packet loss) that were not evaluated in this paper should also be taken into account in a more general NR parametric video quality estimation model.

The current work can be extended in many ways. Firstly, the failure cases (outliers) discovered in the experiment can be exploited to further improve the model. Secondly, a more general model with strong generalization capability is on demand to deal with more challenging service applications (e.g. virtual reality) or more advanced video codecs like HEVC/H.265.

REFERENCES

[1] Pinson, Margaret Helen, L. Janowski, and Z. Papir. "Video Quality Assessment: Subjective testing of entertainment scenes." *IEEE Signal Processing Magazine* 32.1(2015):101-114.

[2] Liu, Yankai, et al. "Review of ITU-T parametric models for compressed video quality estimation." *Signal and Information Processing Association Annual Summit and Conference (APSIPA), 2016 Asia-Pacific*. IEEE, 2016.

[3] ITU-T Recommendation G.1070: *Opinion model for videotelephony applications*, 2012.

[4] ITU-T Recommendation P.1201: *Parametric non-intrusive assessment of audiovisual media streaming quality*, 2012.

[5] ITU-T Recommendation P.1202: *Parametric non-intrusive bitstream assessment of video media streaming quality*, 2012 *assessment of video media streaming quality*, 2012.

[6] ITU-T Recommendation J.343.1: *Hybrid-NRe objective perceptual video quality measurement for HDTV and multimedia IP-based video services in the presence of encrypted bitstream data*, 2014.

[7] Joskowicz, Jose, Rafael Sotelo, and J. Carlos Lopez Ardao. "Towards a general parametric model for perceptual video quality estimation." *IEEE Transactions on Broadcasting* 59.4 (2013): 569-579.

[8] ITU-T Recommendation P.1203: *Parametric bitstream-based quality assessment of progressive download and adaptive audiovisual streaming services over reliable transport,* 2016.

[9] ITU-T Recommendation P.1201.1: *Parametric non-intrusive assessment of audiovisual media streaming quality – lower resolution application area*, 2012.

[10] ITU-T Recommendation P.1201.2: *Parametric non-intrusive assessment of audiovisual media streaming quality – higher resolution application area*, 2012

[11] S. Argyropoulos, P. List, M. N. Garcia, B. Feiten, M. Pettersson and A. Raake, "Scene change detection in encrypted video bit streams," *2013 IEEE International Conference on Image Processing*, Melbourne, VIC, 2013, pp. 2529-2533.

[12] ITU-T Recommendation P.1203.1: *Parametric bitstream-based quality assessment of progressive download and adaptive audiovisual streaming services over reliable transport – Video quality estimation module*, 2016.

[13] Yamagishi, Kazuhisa, T. Kawano, and T. Hayashi. "*Hybrid Video-Quality-Estimation Model for IPTV Services.*" *Global Telecommunications Conference, 2009. GLOBECOM* IEEE Xplore, 2010:1-5.

[14] M. Ries, C. Crespi, O. Nemethova, and M. Rupp, "Content based video quality estimation for H.264/AVC video streaming," in *Proc. WCNC*, 2007, pp. 2668–2673.

[15] J. Joskowicz and J. C. L´opez Ardao, "Combining the effects of frame rate, bit rate, display size and video content in a parametric video quality model," in *Proc. LANC*, 2011, pp. 4–11.

[16] Takagi, Motohiro, et al. "*Subjective video quality estimation to determine optimal spatio-temporal resolution.*" *Picture Coding Symposium IEEE*, 2013:422-425.

[17] China Video Service User Experience Standard Working Group. *China Video Service User Experience.* Beijing: *The People's Posts & Telecom Press*, 2017

[18] ITU-T Recommendation P.910: *Subjective video quality assessment methods for multimedia applications,* 2008.

[19] ITU-T Recommendation P.1401: *Methods, metrics and procedures for statistical evaluation, qualification and comparison of objective quality prediction models*, 2012.